\documentclass{ws-procs9x6}

\begin{document}

\title{Single Spin Asymmetries on a transversely polarised proton target at COMPASS}

\author{S. Levorato on behalf of COMPASS collaboration}

\address{Physics Department, INFN Trieste and Trieste University,\\
Trieste, 34100, Italy\\
$^*$E-mail: stefano.levorato@ts.infn.it\\
}

\begin{abstract} 
COMPASS is a running fixed-target experiment at the CERN SPS with a rich physics program focused on nucleon
spin structure and on hadron spectroscopy. One of the main goals of the spin program is the measurement of the
transverse spin effects in semi-inclusive DIS off transversely polarised nucleons.
In the years 2002, 2003 and 2004 data have been taken using a 160 $GeV/c$ naturally polarised $\mu^{+}$ beam and
a deuterium target ($^{6}LiD$) transversely polarised respect to the beam direction.
In 2007 the run year has been devoted to collect data with  a proton ($NH_{3}$) target.
The preliminary results for the Collins and Sivers asymmetries, extracted from the 2007 data with
transverse target polarisation, are presented here. Results are also compared with existing model predictions. 
\end{abstract}

\keywords{Semi-Inclusive Deep Inelastic Scattering; Transversely polarised proton target; Collins asymmetry; Sivers asymmetry; COMPASS.}

\bodymatter

\section{Single Spin asymmetries}
\label{intro}
The cross-section for polarised deep inelastic scattering \cite{crosssection} of leptons off spin $1/2$ 
hadrons can be expressed, at leading twist, as a function of three independent quark
distribution functions: $q(x)$, $\Delta q(x)$ and  $\Delta_{T}q(x)$. The latter, which describes the probability
 of finding a quark with spin parallel to the nucleon spin in a transversely polarised nucleon, 
is chiral-odd and can be measured in transversely polarised nucleon (anti)-nucleon hard scattering 
or in semi-inclusive deep inelastic scattering (SIDIS) on a transversely polarised target.\\ 
In SIDIS $\Delta_{T}q(x)$ can be measured in combination with the chiral odd Collins fragmentation 
function $\Delta _{0}^{T}D^{h}_{q}$, via azimuthal single spin asymmetries (SSA) in single hadron production. 
According to Collins\cite{collins}, the fragmentation of a transversely  polarised quark 
in unpolarised hadrons presents an azimuthal modulation
with respect to the plane defined by the quark momentum and the quark spin.
In SIDIS the hadron yield can be written then as:
 \begin{equation}
N = N_0 \cdot (1+f \cdot P_t \cdot D_{nn} \cdot A_C \cdot \sin(\Phi_C)) 
\label{eq:collfun}
\end{equation}
where $f$ is the target dilution factor, $P_t$ the target polarisation and $ D_{nn}  = (1-y)/(1-y+y^2/2)$ the transverse spin
transfer coefficient from the initial to the struck quark. 
The angle $\Phi_{C}$ is known as ``Collins angle'' and  is conveniently defined in the system where the z-axis is the 
virtual photon direction and the x-z plane is the muon scattering plane. 
In this frame $\Phi_{C}= \Phi_h+\Phi_S-\pi$, where $\Phi_h$ is the hadron azimuthal angle, and
$\phi_S$ the azimuthal angle of the transverse spin of the initial
nucleon. Finally $A_C$ is the Collins asymmetry resulting from the convolution between the  Collins fragmentation function and the transverse spin distribution:
\begin{equation}
A_C= \frac{\sum_q e_q^2 \, \Delta_T q(x) \, \Delta_T^0 D_q^h(z, p_{T}^{h})}
       {\sum_q e_q^2 \,     q(x) \,       D_q^h(z, p_{T}^{h})} 
\label{eq::collasim}
\end{equation}
where $e_q$ is the quark charge, $D_q^h(z, p_{T}^{h})$ is the unpolarised fragmentation function, $z= E_h /(E_{l}-E_{l'})$ is the fraction of available energy carried by the hadron,
and $p_{T}^{h}$ is the hadron transverse momentum with respect to the virtual photon direction.
As is clear from eq.\ref{eq:collfun}, the Collins asymmetry  $A_C$ is  revealed as a $\sin \Phi_{ C}$ modulation
in the number of produced hadrons.\\
A second source of azimuthal asymmetry is related to the Sivers effect~\cite{sivers},
arising from a possible coupling 
of the intrinsic transverse momentum $\vec k_T $
of unpolarised quarks to the spin of a transversely polarised nucleon. 
In this case the number of  produced hadrons can be written as:
\begin{equation}
N = N_0 \cdot (1+f  \cdot P_t  \cdot A_S  \cdot \sin(\Phi_S)) 
\end{equation}
where the Sivers angle $\Phi_S$ is defined as $\Phi_h - \Phi_{s}$,
and the asymmetry $A_S$ probes the so called Sivers distribution function  $\Delta_0^T q$:
\begin{equation}
A_{S}  =  \frac {\sum_q e_q^2 \, \Delta_0^T q (x,\vec k_T) \, D^h_q  (z)}
{\sum_q e_q^2 \, q (x) \, D_q^h(z)} .
\end{equation}
In this case the asymmetry $A_S$ is revealed as a $\sin \Phi_{ S}$ modulation 
in the number of produced hadrons.\\ It has to be noted here that 
since the Collins and the Sivers angles are independent~\cite{ind}, 
it is possible to measure from the same data both the Collins and the Sivers asymmetries.\\

\section{Data sample and analysis}
\begin{figure}
\begin{center}
\psfig{file=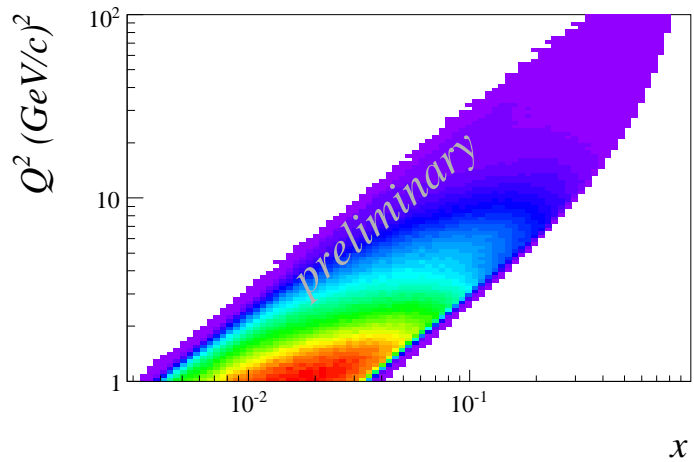,width=10 cm}
\caption{Scatter plot of $Q^2$ as function of \textit{x} Bjorken for the events after all the kinematic cuts. Most of the statistics is concentrated at low \textit{x} Bjorken values.}
\label{fig::q_x}
\end{center}
\end{figure}

The Collins and the Sivers asymmetries have been measured by COMPASS from 2002 to 2004 on a transversely polarised deuterium target; both the asymmetries turned out to be small, compatible with zero\cite{compass,ageev,alekseev}. 
In 2007 COMPASS took data with a proton target ($NH_3$) and a 50\% sharing between longitudinal and transverse target configuration,
accumulating $\sim 40\cdot 10^{12}$ and $\sim 42\cdot 10^{12}$ $\mu$ 
on tape respectively.\\ A new target with a larger diameter (4 cm) and three cells (upstream, central, downstream, 30, 60, and 30 cm long respectively) have been used: the tree cells setup is meant to reduce the difference in acceptance between the two polarisations. The upstream and the downstream cells are polarised along the same direction, which is opposite to the central cell. The Ammonia material is characterized by a dilution factor f $\sim$ 0.15 and a very  high polarisation: $\sim$ 90\%. As in 2006 the target is inside the new large acceptance Superconducting magnet.
The transverse run has been divided in 12 "periods",  each of them 
corresponding to about 5 full days of data taking. Consecutive periods correspond to opposite polarisation of each of the tree target 
cells. Asymmetries are extracted using at the same time the informations coming from cells in two consecutive periods with opposite configuration.\\
Almost all the data collected  in the transverse polarisation
configuration of the target have been processed a first time. For the results 
presented here about 20\% of the whole collected data have been used. These data have been selected requiring a good stability of the spectrometer and of the reconstruction between consecutive periods.
In the  analysis the events are considered only if one primary vertex is found in the target region. To select DIS
events the photon virtuality $Q^{2}$ is taken greater than 1 (GeV/$c$)$^2$, $y$ between 0.1 and 0.9, and the invariant
mass of the final hadronic state $W>5$ GeV/$c^{2}$. The hadron sample on which the asymmetries are computed consists of all the charged hadrons coming
from the reaction vertex with $p^{h}_T > 0.1$ GeV/$c$ and $z$ $>$ 0.2. In table \ref{tab:final_stat} the final statistics entering the
asymmetries extraction is given for the 6 used periods, separately for positive and negative hadrons.
\begin{table}
\tbl{Number of hadrons used for this analysis.}
{\begin{tabular}{@{}lcc@{}}\toprule
Period & Positive hadrons:  & Negative hadrons \\
\colrule
W39/W40   & 2742704 & 2149343 \\
W41/W42a & 2199513 & 1752684 \\
W42b/W42 &  761014 &  603061 \\
\hline
Total: & 5703231 & 4505088\\
\botrule
\end{tabular}}
\label{tab:final_stat}
\end{table}
Figure \ref{fig::q_x} illustrates the $Q^2$-\textit{x} Bjorken phase-space covered by the COMPASS experiment after the $Q^{2}>1$ selection. 
As it can be seen the high energy of the muon beam allow to reach the $10^{-3}$ region of \textit{x}, in the DIS regime, moreover most of the statistic is at low \textit{x} values: 0.008 $\div$ 0.02.\\     
A number of systematic studies have been performed in order to determine the systematic errors. Extensive tests both to measure false asymmetries and to investigate 
the stability of the physics results, in each data taking period, were done combining cells with the same polarisation and using  different splitting combination of the target cells (top-bottom, left-right). The results obtained with different asymmetries extraction methods have been compared too.\\
From all these tests the systematic errors have been estimated to be $0.3\cdot\sigma_{stat}$  for the Collins asymmetries and $0.5\cdot\sigma_{stat}$ for the Sivers asymmetries.

\subsection{Results}
\label{result}
\begin{figure}
\begin{center}
\psfig{file=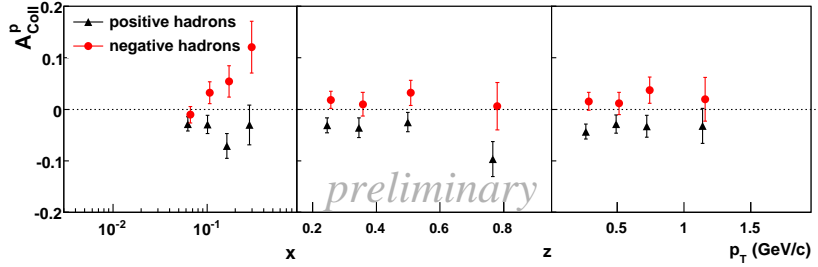,width=11 cm}
\caption{Collins asymmetries for positive unidentified hadrons (triangles) and for negative unidentified hadrons (circles) with the cut $x>0.05$  as function of $x$, $p^h_T$, and $z$.}
\label{collinsprotonhrange}
\end{center}
\end{figure}

The Collins and Sivers asymmetries were evaluated as a function of $x$, $p^h_T$, and $z$ dividing the corresponding kinematical range in 
bins (with variable width, in order to have  a comparable statistics in each of them), and integrating over the other two
variables. In total, the asymmetries were evaluated in 9  $x$-bins, 9  $p^h_T$-bins, and 8 $z$-bins.
The method used for extraction is based on a two dimensional (8 times 8) binning in $\Phi_h$ and $\Phi_S$. 
Combining the information of the opposite target polarisation cells a non-linear system of equations for the cross section amplitude modulations can be written, and the extraction of the amplitudes is then obtained via maximum likelihood method. The results have been checked with several other statistical methods described in Ref. 6.
\begin{figure}
\begin{center}
\psfig{file=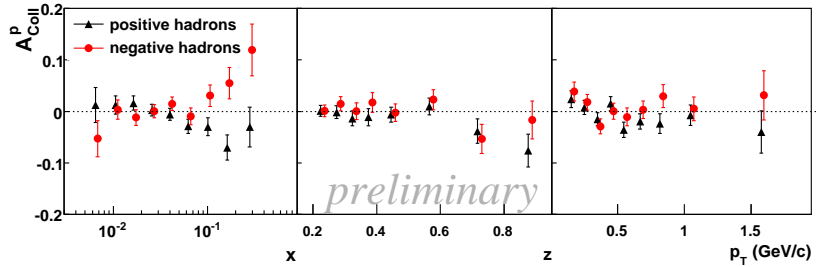,width=11 cm}
\caption{Collins asymmetries for positive unidentified hadrons (triangles) and for negative unidentified hadrons (circles) as function of $x$, $p^h_T$, and $z$. }
\label{collinsproton}
\end{center}
\end{figure}
In figure \ref{collinsproton} the preliminary results for the Collins asymmetries as function of $x$, $p^h_T$, and $z$  are shown both for positive and for negative unidentified hadrons. The asymmetry is small, basically statistically compatible with zero up to $x=0.05$ while a signal is visible in the last points: 
the asymmetry then increases in module up to 10\% and with oppoiste sign for positive and negative hadrons. For  $p^h_T$ and $z$  the asymmetry amplitude is compatible with zero due to the fact that most of the statistical sample is in the low \textit{x} region. In fact requiring  $x>0.05$ the asymmetry signal becomes more evident both in $p^h_T$ and $z$ bins. This can
be seen in fig. \ref{collinsprotonhrange}, showing Collins asymmetry for the data in the valence region, i.e. with $x>0.05$. It is clear from the plots that there is not an appreciable $z$ or $p^h_T$ dependence.
 
\begin{figure}
\begin{center}
\psfig{file=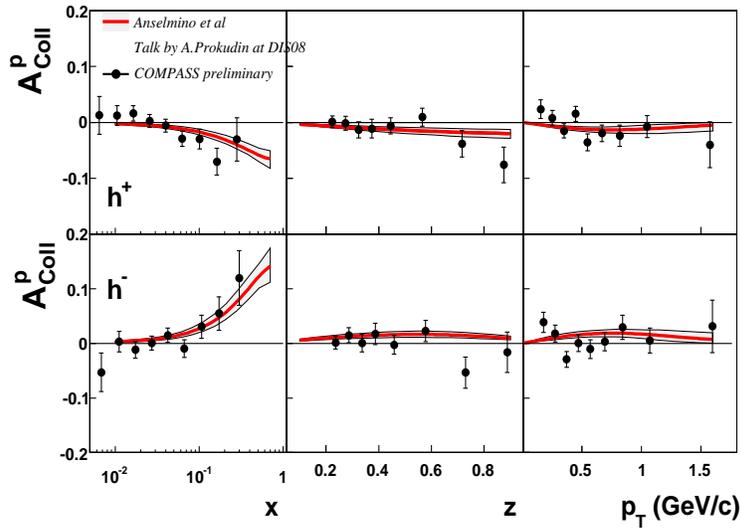,height= 7.5cm,width=10.5 cm}
\caption{Compass preliminary Collins asymmetries on proton with the latest prediction 
of Anselmino et \textit{al}, (A. Prokudin at DIS 08). }
\label{predictioncollins}
\end{center}
\end{figure}
\begin{figure}
\begin{center}
\psfig{file=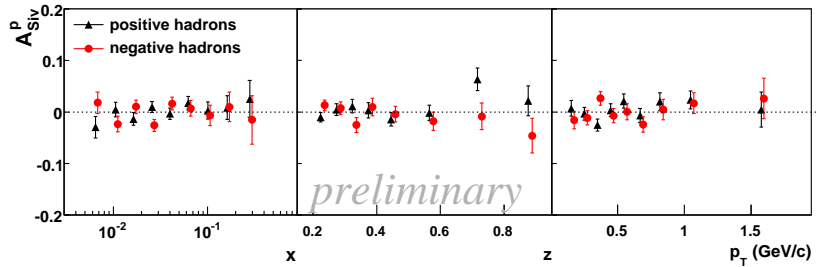,width=11 cm}
\caption{Sivers asymmetries for positive unidentified hadrons (black) and for negative unidentified hadrons (circles) as function of $x$, $p^h_T$, and $z$.}
\label{siversproton}
\end{center}
\end{figure}

Figure \ref{predictioncollins} shows the theoretical expectations of the Collins asymmetries in the COMPASS kinematical range\cite{prokudin} both for positive and negative hadrons. The predictions are obtained from a global analysis of the HERMES\cite{Diefenthaler:2007rj} data, COMPASS \cite{compassspectro} deuterium data and the BELLE\cite{belle} data. The good agreement for all the different kinematic variables is manifested in the figure. 
In figure \ref{siversproton} the preliminary results of the Sivers asymmetries are shown as function of  $x$, $p^h_T$ and $z$. At variance with $A_{C}$, the Sivers asymmetry is small and statistically compatible with zero for both positive and negative hadrons over all the measured \textit{x} range. 
The result for positive hadrons is at variance from what has been measured by HERMES \cite{Diefenthaler:2007rj}, and the two data samples are marginally compatible. A possible explanation of this result may come from the very different kinematic range of the two experiments.
 
\begin{figure}[ht]
\begin{center}
\psfig{file=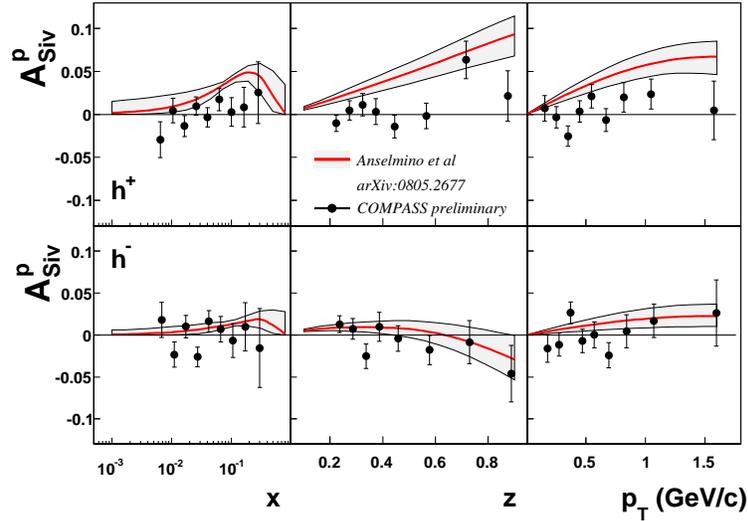,height=7.5cm,width=10.5 cm}
\caption{COMPASS Sivers asymmetry on proton for positive and negative hadrons,
with the latest prediction of Anselmino et \textit{al}.}
\label{predictionsivers}
\end{center}
\end{figure}

Figure \ref{predictionsivers} compares the COMPASS Sivers asymmetry on proton data for positive and negative hadrons with the latest prediction of Anselmino et \textit{al}\cite{Anselmino:2008sg}.. For positive hadrons the agreement is not satisfactory. As in the Collins case  the predictions  are driven by the HERMES data.\\
Other theoretical models as for example the one proposed by S. Arnold et \emph{al}\cite{efremov} show the same disagreement for the Sivers results. 

\section{Summary}
Preliminary results of Collins and Sivers asymmetries for 2007 COMPASS proton data have been presented. Collins asymmetries for positive and negative hadrons are different from zero and of opposite sign for the two charges and agree with the previous HERMES results. In the Sivers case the measured asymmetries are compatible with zero, within the present statistics, both for positive and negative hadrons, at variance with the HERMES result.\\

\bibliographystyle{ws-procs9x6}
\bibliography{ws-pro-sample}

\end{document}